\documentclass[twocolumn,showpacs,amsmath,amssymb,prl]{revtex4}

\usepackage{graphicx}

\newcommand{\be}{\begin{equation}}
\newcommand{\ee}{\end{equation}}
\newcommand{\bfig}{\begin{figure}}
\newcommand{\efig}{\end{figure}}

\begin{document}
\title{Pressure-Induced Electronic Transition in Black Phosphorus
}
\author{Z. J. Xiang$^1$, G. J. Ye$^1$, C. Shang$^1$, B. Lei$^1$, N. Z. Wang$^1$, K. S. Yang$^3$, D. Y. Liu$^3$, F. B. Meng$^1$, X. G. Luo$^{1,6}$, L. J. Zou$^3$, Z. Sun$^{4,6}$}
\author{Y. Zhang$^{5,6}$}
 \email{zhyb@fudan.edu.cn}
\author{X. H. Chen$^{1,2,6}$}
 \email{chenxh@ustc.edu.cn}

\affiliation{
$^1$ Hefei National Laboratory for Physical Sciences at Microscale and Department of Physics, University of Science and Technology of China, Hefei, Anhui 230026, China and Key Laboratory of Strongly-coupled Quantum Matter Physics, Chinese Academy of Sciences, Hefei, Anhui 230026, China\\
$^2$ High Magnetic Field Laboratory, Chinese Academy of Sciences, Hefei, Anhui 230031, China\\
$^3$ Key Laboratory of Materials Physics, Institute of Solid State Physics, Chinese Academy of Sciences, Hefei, Anhui 230031, China\\
$^4$ National Synchrotron Radiation Laboratory, University of Science and Technology of China, Hefei, Anhui 230026, China\\
$^5$ State Key Laboratory of Surface Physics and Department of Physics, Fudan University, Shanghai 200433, China\\
$^6$ Collaborative Innovation Center of Advanced Microstructures, Nanjing 210093, China
}

\date{\today}
\begin{abstract}
In a semimetal, both electrons and holes contribute to the density of states at the Fermi level. The small band overlaps and multiband effects engender novel electronic properties. We show that a moderate hydrostatic pressure effectively suppresses the band gap in the elemental semiconductor black phosphorus. An electronic topological transition takes place at approximately 1.2 GPa, above which black phosphorus evolves into a semimetal state that is characterized by a colossal positive magnetoresistance and a non-linear field dependence of Hall resistivity. The Shubnikov-de Haas oscillations detected in magnetic field reveal the complex Fermi surface topology of the semimetallic phase. Specially, we find a non-trivial Berry's phase in one Fermi surface that emerges in the semimetal state, as evidence of a Dirac-like dispersion. The observed semimetallic behavior greatly enriches the material property of black phosphorus, and sets the stage for the exploration of novel electronic states in this material.
\end{abstract}
\maketitle                   
The electronic band gap is an intrinsic feature of a semiconductor that largely determines the semiconductor's electronic and optical properties. In black phosphorus, which recently emerged as a unique layered semiconductor that exhibits high carrier mobility \cite{Likai,Qiao}, the puckered honeycomb lattice in each P atom layer \cite{Morita} dictates that the direct band gap is mainly determined by the out-of-plane $\emph{p}_z$-like orbital of phosphorus and further reduced by the interlayer coupling in the bulk crystal \cite{Takao,Asahina}. Thus, a slight change in the height of the puckered layers strongly modulates the direct band gap of black phosphorus \cite{Rodin}.  An applied hydrostatic pressure can close the band gap \cite{Morita,Keyes}, and drive a structural transition above 4 GPa \cite{Okajima}. The effectiveness of such pressure-induced band modification and the theoretically suggested linear band dispersion along the armchair direction in black phosphorus \cite{Fei} enable exploring exotic electronic phases in this material. In this study, we discovered that a moderate pressure (substantially below the critical pressure of the structural transition) not only closes the band gap, but also induces an electronic topological transition (also known as Lifshitz transition \cite{Lifshitz}) that fundamentally modifies the material property of black phosphorus. Specifically, black phosphorus evolves into a semimetal above the Lifshitz transition. Consequently, we observed a colossal magnetoresistance up to approximately 80000$\%$ at a magnetic field of 9 T, which was possibly related to the nearly compensated electrons and holes in this semimetal state. Multiple electronic bands emerged from the Lifshitz transition, and produced quantum oscillations when the sample was subjected to external magnet fields. Analysis of the oscillation phases revealed that one of the emergent bands carried a non-trivial Berry's phase, a prominent feature of a Dirac-like dispersion.

We used single crystals of black phosphorus that were prepared using the high-pressure synthesis technique (details provided in Supplemental Material \cite{SM}). The sample was a $p$-type semiconductor at ambient pressure (Supplemental Material \cite{SM}). The typical sizes of single crystal samples were 2.5$\times$0.7$\times$0.1 mm$^{3}$, and the electrical contacts were arranged in a conventional Hall-bar configuration. We applied hydrostatic pressure by using a self-clamped BeCu piston-cylinder cell that produces pressures up to 2.4 GPa. The magneto-transport measurements were performed in a Quantum Design PPMS-9.

The application of hydrostatic pressure drastically changes the transport properties of black phosphorus, and induces a semiconductor-to-metal transition. Figure 1 displays the temperature dependence of the longitudinal resistivity $\rho_{xx}$ of a black phosphorus sample under varying pressures. The low-temperature resistive divergence was gradually suppressed by the applied pressure and the upturn feature completely vanished at \emph{P} = 1.25 GPa. At \emph{P} = 1.38 GPa, the high-temperature non-metallic behavior of $\rho_{xx}$($\emph{T}$) was not detectable up to room temperature. Metallic resistivity in the entire temperature range from 2 to 300 K was observed at higher pressures. Such behaviors can generally be captured by the ratio between the resistivity at room temperature and that at the lowest temperature, also referred to as the residual resistivity ratio (RRR). As shown in the inset of Fig. 1, a prominent kink in the RRR indicates a semiconductor-to-metal transition occurring at a critical pressure of $\emph{P}_c$ $\simeq$ 1.2 GPa. Intriguingly, such a critical pressure coincides with a drastic enhancement of the sample magnetoresistance (MR) when the magnetic field is applied normal to the P atom plane (along the $c$-axis). The concurrent sharp enhancement of RRR and MR points to a profound change in the electronic structure of black phosphorus at $\emph{P}$ = $\emph{P}_c$.

\begin{figure}[h]
\centering
\includegraphics[width=0.48\textwidth]{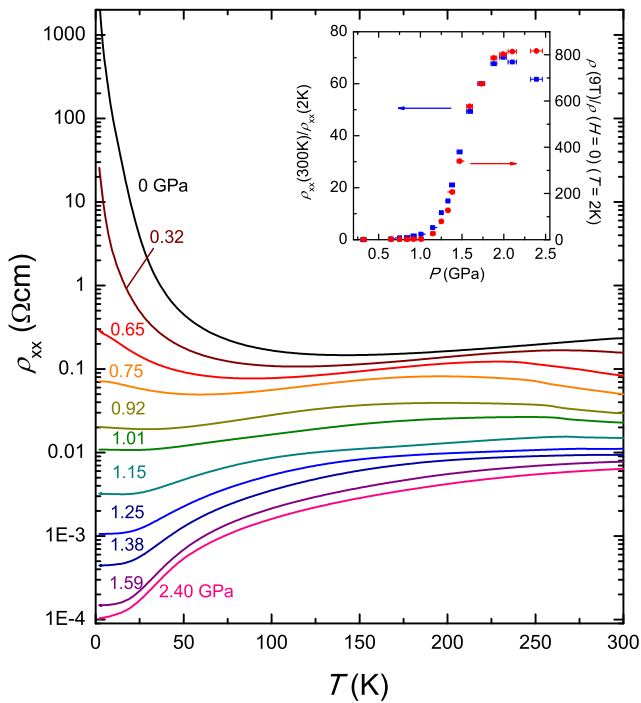}
\caption{(color online). Longitudinal resistivity $\rho_{xx}$ of black phosphorus versus \emph{T} at various hydrostatic  pressures. Numbers  beside the curves represent applied pressures in gigapascals. Inset: The value of the residual resistivity ratio $\rho_{xx}$(300 K)/$\rho_{xx}$(2 K) and the transverse MR ([($\rho_{xx}$(9T)-$\rho_{xx}$(\emph{H}=0)]/$\rho_{xx}$(\emph{H}=0)) with \emph{H} $\parallel$ c at $\emph{T}$ = 2 K as a function of pressure.
}
\label{FIG. 1}
\end{figure}

Previous studies have confirmed that the orthorhombic ($A$17) structure of black phosphorus is preserved up to 4.5-5 GPa without any discontinuity in the structural parameters \cite{Akahama1,Kikegawa}, meaning that the transition we observed was an isostructural electronic transition. Moreover, the slope change of the pressure dependent resistivity at the critical pressure (Supplemental Material \cite{SM}) was consistent with an electronic topological transition \cite{Blanter,JZhang}. The field dependence of the Hall resistivity provided strong evidence of co-existing electron and hole bands above $\emph{P}_c$, as shown in Fig. 2(a), a hallmark of multiband conduction, became apparent at $\emph{P}$ $>$ 1.20 GPa and extended to high field under higher pressure. Furthermore, the sign of $\rho_{xy}$ reversed from negative (electron-type) to positive (hole-type) at $\emph{H}$ $\simeq$ 0.11 T at 1.20 GPa, and the sign reversal field shifted upward as the pressure increased (Fig. 2(a), inset). The Hall resistivity changes its sign in magnetic field when one of the two types of carriers (electrons and holes) has larger carrier population and the other shows higher mobility, as observed in graphite \cite{Soule} and NbSb$_2$ \cite{Petrovic}. Therefore, at $\emph{P}_c$ $\approx$ 1.20 GPa, black phosphorus is likely to enter a semimetal phase in which both electrons and holes are present at the Fermi level. Based on all of these findings, we identify the semiconductor-to-semimetal transition in black phosphorus as the topological Lifshitz transition of the band structure: The applied pressure pushed the conduction and valence bands toward each other, and eventually produced a semimetal state when the bands crossed the Fermi level successively.

A notable consequence of this electronic transition is the observation of a colossal MR in the semimetallic phase of black phosphorus. When the pressure increased from $\emph{P}_c$ $\approx$ 1.2 GPa up to approximately 2.0 GPa, the positive MR was enhanced by a factor of $\sim$ 40 and reached the exceedingly high value of $\sim$80000$\%$ in a magnetic field of 9 T, with no indication of saturation [Fig. 2(b)]. A similar non-saturating large positive MR has been reported in a number of semimetals \cite{Soule,Alers,Ali}, and attributed to electron-hole compensation \cite{Ali,Du}. The colossal MR was also observed with $\emph{H}$ parallel to the basal plane (Supplemental Material \cite{SM}) in spite of the expected anisotropy \cite{Morita,Strutz}.

\begin{figure}[h]
\centering
\includegraphics[width=0.48\textwidth]{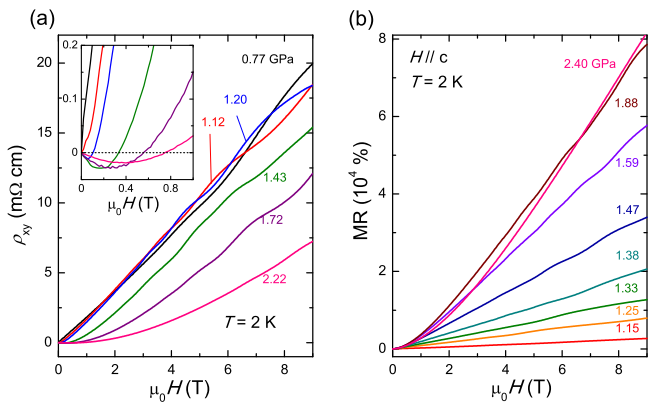}
\caption{(color online). (a) Hall resistivity $\rho_{xy}$ of a black phosphorus single crystal as a function of the magnetic field at various pressures at $\emph{T}$ = 2 K. Inset: Low-field sign reversal of $\rho_{xy}$ at $\emph{P}$ $\geq$ 1.20 GPa. The Hall data were antisymmetrized with respect to positive and negative magnetic fields to remove the $\rho_{xx}$ component. (b) Field dependence of transverse MR for $\emph{P}$ $>$ 1.1 GPa with \emph{H} $\parallel$ c, and current $\emph{I}$ applied in the basal plane, at $\emph{T}$ = 2 K.}
\label{FIG. 2}
\end{figure}

Quantum oscillations in magnetic field serve as a powerful tool for probing the Fermi surface (FS) topology across the pressure-induced Lifshitz transition. Shubnikov de Haas (SdH) oscillations emerged even before the low-temperature upturn of resistance was completely suppressed. Hints of SdH extrema on MR curves were already discernible at $\emph{P}$ = 0.65 GPa (arrows in Supplemental Material Fig. S2 \cite{SM}) and the oscillation frequency could be resolved above 0.75 GPa. Oscillation patterns at various pressures are plotted in Fig. 3(a). Clear peaks manifested in fast Fourier transform (FFT) spectra [Fig. 3(b)]. The oscillation frequencies are remarkably low, reflecting the small sizes of Fermi pockets. At pressures below 1.25 GPa, a magnetic field of 9 T can force all the electrons into the lowest Landau level. According to the Lifshitz-Onsager quantization rule, the quantum oscillation frequency ($\emph{F}$) is proportional to the extremal cross-sectional area $S_F$ of the FS \cite{Shoenberg}, $\emph{F}$ =  ($\hbar$/2$\pi$e)$\emph{S}_F$ . The lowest frequency measured with \emph{H} $\parallel$ c was $\emph{F}$ = 4.0 T at $\emph{P}$ = 0.92 GPa, corresponding to a cross-sectional area $\emph{S}_F$ = 0.038 nm$^{-2}$, which was only about 0.014$\%$ of the area of Brillouin zone in (001) plane. More than one oscillation frequency could be resolved at $\emph{P}$ $\geq$ 1.25 GPa (Fig. 3(b)), indicating the existence of multiple FS extrema. Notably, the starting point of the multiple-frequency region was consistent with the onset pressure of the semimetal phase. The emergence of a new SdH branch strongly suggests that the band topology is modified at this electronic transition: Combining this phenomenon with the Hall result, there is a reasonable assumption that an electron FS appears at the critical pressure.

In the FFT spectra of the semimetal phase, two sets of spectrally prominent features, $\alpha$ and $\beta$, were identified based on their pressure dependence above 1.38 GPa [Fig. 3(b) and (c)]. An Expansion of the FSs under pressure could be observed in both branches. At high pressures, branch $\alpha$ developed into the strongest feature, whereas branch $\beta$ became comparably weak. We note that multiple peaks appear in the FFT spectra between 1.25 GPa and 1.38 GPa, suggesting the existence of additional FS branches. The quantitative identification of such branches is however difficult in the data set currently available.

Analyzing the temperature dependence of the oscillation amplitude based on the Lifshitz-Kosevich (LK) formula \cite{Shoenberg} yielded small carrier cyclotron masses: the cyclotron mass ratio $\emph{m}$* = $\emph{m}$/m$_e$ (where m$_e$ is the free electron mass) varied from 0.043 to 0.136 in the pressure range from 0.75 to 1.59 GPa and its pressure dependence mostly matched that of the corresponding oscillation frequencies [Fig. 3(d)] (for details see Supplemental Material \cite{SM}). The extracted carrier masses were considerably smaller than that reported for ambient pressure \cite{Morita,Likai2,mass} and this notable mass reduction reflected a drastically modified electronic structure under pressure. SdH oscillations were also clearly resolved when a magnetic field was applied in the basal plane, and small Fermi pockets and small cyclotron masses were again observed (Supplemental Material Fig. S7 \cite{SM}). The oscillations observed for $\emph{H}$ $\perp$ c confirmed that at least one of the FSs was three dimensional (3D) in nature with close orbit perpendicular to the basal plane.

\begin{figure}[h]
\centering
\includegraphics[width=0.48\textwidth]{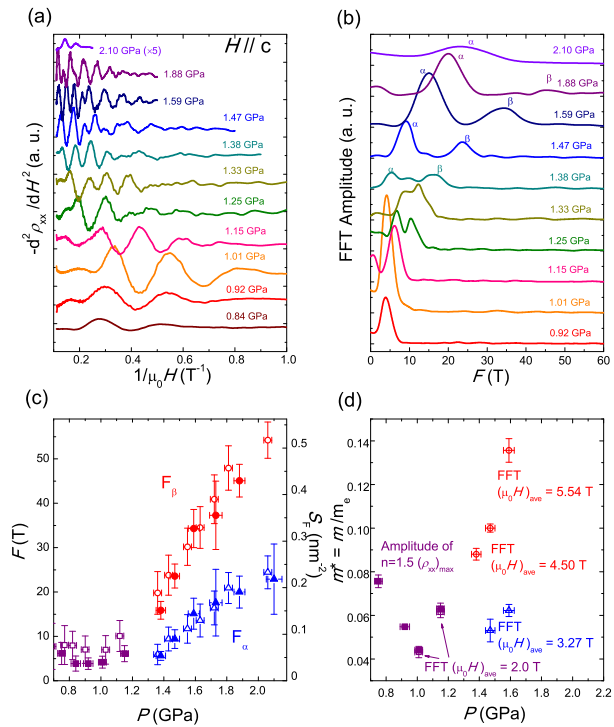}
\caption{(color online). (a) Negative second derivative of resistivity with respect to the magnetic field, -d$^2$$\rho_{xx}$/d$\emph{H}$$^2$, at varying pressures. Magnetic field $\emph{H}$ was applied along the $c$ direction. Data were collected for a black phosphorus single crystal labeled sample $A$. (b) The FFT spectra of -d$^2$$\rho_{xx}$/d$\emph{H}$$^2$ of sample $A$. The labels $\alpha$ and $\beta$ are assigned to the resolved FFT peaks. (c) The oscillation frequencies and corresponding cross-sectional areas of Fermi surfaces of sample $A$ (solid symbols) and sample $B$ (open symbols), plotted as a function of $\emph{P}$. The uncertainty is determined by the half width at half-height (HWHH) of the FFT peaks. The raw data obtained for sample $B$ is shown in Supplemental Material Fig. S6 \cite{SM}. (d) Pressure dependence of the cyclotron mass ratio $\emph{m}$* = $\emph{m}$/m$_e$. $\emph{m}$* was calculated based on the LK formula fitting of the $\emph{T}$ dependence of the normalized amplitude of the oscillatory component $\Delta$$\rho_{xx}$/$\rho_{xx}$ (solid symbols) or the normalized FFT amplitude (open symbols). The average inverse fields in the FFT interval are noted beside data points.}
\label{FIG. 3}
\end{figure}

We now turn to the phases of quantum oscillations in the pressure-induced semimetal state. The SdH oscillation can be semi-classically described as follows \cite{Shoenberg,Roth,Xiong}:
\begin{equation}
\Delta\sigma_{xx} \varpropto cos[2\pi(\frac{F}{B}-\gamma+\delta)],
\label{SdH}
\end{equation}
where $\sigma_{xx}$ is the longitudinal conductivity, $\emph{F}$ the oscillation frequency, $\gamma$ the Onsager phase factor and $\delta$ an additional phase shift that takes a value between -1/8 and +1/8 depending on the degree of three-dimensionality of the FS. The Onsager phase $\gamma$ relates to the Berry's phase $\phi_B$ through $\gamma$ = 1/2-$\phi_B$/2$\pi$ \cite{Taskin,Mikitik}. In an ordinary electron system that exhibits parabolic energy dispersion, $\phi_B$ is zero and therefore $\gamma$ = 1/2. By contrast, in a massless Dirac electron system showing linear energy dispersion, a non-trivial Berry's phase $\phi_B$ = $\pi$ exists, yielding $\gamma$ = 0 \cite{Mikitik,YZhang}. For the frequency component $\alpha$ which is spectrally dominate above 1.6 GPa [Fig. 4(a)], we could extract the value of $\gamma$ from the Landau index plots \cite{Landau fan}. The linear extrapolation of 1/$B_n$ versus the integer $\emph{n}$ plot is presented in Fig. 4(b), and the inset summarizes the -$\gamma$+$\delta$ values for the $\alpha$ pocket, all of which are close to 0 with offsets lower than 0.07. Since black phosphorus is a quasi-two-dimensional system, the absolute value of the phase shift $\delta$ can be smaller than that in the 3D limit (1/8) \cite{Murakawa}. Thus the result of intercept analysis indicated a phase factor $\gamma$ $\simeq$ 0. The zero Onsager phase corresponded to a non-trivial Berry's phase $\phi_B$ $\simeq$ $\pi$, providing evidence of the linear band dispersion \cite{YZhang} for pocket $\alpha$ (analysis of the intercept for other pockets is less conclusive owing to the relatively weak signal). Our results suggested that black phosphorus could be a new candidate for the elemental semimetals that possibly host exotic electronic states, besides graphite \cite{Zhou,GLi} and bismuth \cite{Lu}.

\begin{figure}[h]
\centering
\includegraphics[width=0.48\textwidth]{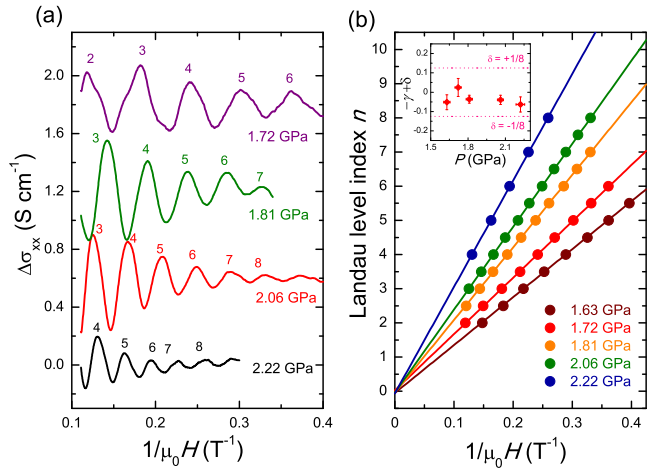}
\caption{(color online). (a) Oscillation patterns of the magneto-conductivity $\sigma_{xx}$(\emph{H}) of pocket $\alpha$, obtained by subtracting a smooth background from  $\sigma_{xx}$=$\rho_{xx}$/($\rho^2_{xx}$+$\rho^2_{xy}$), at various pressures with \emph{H} $\parallel$ c. Integer indices are marked at the maxima of $\Delta$$\sigma_{xx}$. (b) Landau index \emph{n} versus 1/$\mu_0$\emph{H} for the SdH oscillation of pocket $\alpha$. The linear extrapolations of the indices are indicated by solid lines. Inset: Intercepts of the linear extrapolations on \emph{n}-axis, which equal to the total SdH phase -$\gamma$+$\delta$. The correction terms $\delta$  in the 3D limit, $\delta$ = -1/8 (FS maxima) and $\delta$ = +1/8 (FS minima), are denoted by dashed lines for the phase of Dirac electrons ($\gamma$ = 0). Error bars represent the uncertainties of linear fitting.
}
\label{FIG. 4}
\end{figure}

The Lifshitz transition occurring at $\emph{P}_c$  $\simeq$ 1.2 GPa stems from the lattice distortions of black phosphorus under pressure. Although the covalent bonds between phosphorus atoms resist against compression, the change in the bond angle and the weak van der Waals interlayer interactions result in large compressibility in $a$ and $c$ direction, respectively \cite{Morita}. Therefore, black phosphorus is extremely soft and exhibits an extraordinarily pronounced pressure-induced band transformation for an inorganic semiconductor. Specifically, a recent first-principles calculation showed that the $\emph{p}_z$ orbitals of adjacent atoms in distinct sub-lattices of the puckered lattice of black phosphorus gradually overlap under pressure \cite{Gong}. This orbital overlap drives the band crossover near the $Z$ point in Brillouin zone, develops multiple Dirac cones and produces nearly compensated electron and hole pockets. In their band structure calculation there are three types of Fermi pockets in the semimetal phase and two of them are close to Dirac points \cite{Gong}. This prediction is partially supported by our observations, although the third FS and a second Dirac-like band remain to be identified. Similar trend of band structure evolution in black phosphorus has been suggested in previous theoretical studies \cite{Rodin,Fei}.

The qualitative consistence of calculation works and our experimental results strongly suggests a plausible scenario: black phosphorus evolves from a narrow-gap semiconductor to a 3D Dirac semimetal under external pressure, as a result of the formation of band crossover with linear dispersions. The Lifshitz transition we observed can be related to the theoretically predicted splitting of an initial four-fold degenerate Dirac point at the $Z$ point \cite{Gong}. An alternative explanation is that the van Hove singularity of the upper band passes the Fermi level at the critical pressure \cite{Enderlein}, resulting in a new electron pocket. We note that one FS emerges below $\emph{P}_c$ as an SdH frequency was already resolved at $\sim$ 0.7 GPa. Since the black phosphorus samples are slightly $p$-doped and the Fermi level is closer to the top of valence band, it is possible that  before the complete closure of the band gap, the Fermi level crosses the edge of valence band and a small hole pocket is formed. Therefore, when pressure is applied, black phosphorus may undergoes more than one electronic transition. Additional investigations are required for determining the exact pressure at which the band gap closes and for clarifying the details of band structure evolution under pressure.

In summary, we observed a phase transition that introduces a typical semimetal phase in the direct gap semiconductor black phosphorus under a moderate hydrostatic pressure of $\emph{P}_c$ $\simeq$ 1.2 GPa, which we identified as a topological Lifshitz transition of the electronic band structure. The semimetal phase was characterized by the colossal MR, the coexistence of electron and hole pockets, low carrier densities, and small cyclotron masses. A non-trivial Berry's phase was detected for one of the pockets, indicating a potentially peculiar band topology. Our results demonstrate that pressure, apart from layer number, serves as an effective parameter of tuning the electronic structure of black phosphorus. The dramatic response of black phosphorus to modest pressure reflects this material¡¯s capacity for sensitively detecting pressure or strain. Moreover, the evidence of linear dispersion in bulk black phosphorus under pressure points to a potential two-dimensional Dirac fermion system in phosphorene, a monolayer of black phosphorus.

\vspace{2ex}We acknowledge helpful discussions with Y. Luo, S. Q. Shen, L. P. He and S. Y. Li. This work is supported by National Natural Science Foundation of China (NSFC), the "Strategic Priority Research Program (B)" of the Chinese Academy of Sciences, the National Basic Research Program of China (973 Program).

%

%

\begin{thebibliography}{99}
\bibitem{mass}
Without applied pressure, the cyclotron mass for holes in $xy$-plane is $m$ = ($m_x$$m_y$)$^{1/2}$ = 0.30 m$_e$ deduced from the value of effective masses ($m_x$ = 0.12 m$_e$ and $m_y$ = 0.72 m$_e$)(see Ref. \cite{Likai2}), more than two times larger than the largest cyclotron mass we got: $m$ = 0.136 m$_e$ for pocket $\beta$ at 1.59 GPa.
\end{thebibliography}
\end{document}